\documentclass{cimento}

\title{Guido Altarelli and the evolution of QCD\thanks{Durham preprint IPPP/16/77}}
\author{R. Keith Ellis%\from{ins:x}
}
\instlist{\inst{} 
Institute for Particle Physics Phenomenology,
Department of Physics, Durham University, Durham DH1 3LE, United Kingdom}
\begin{document}
\maketitle

\begin{abstract}
I describe the contributions of Guido Altarelli to the development of Quantum Chromodynamics 
from the discovery of
asymptotic freedom until the end of the S$p\bar{p}$S collider era, 1973-1985.
\end{abstract}

\section{Introduction}
I have been asked to write an appreciation of Guido Altarelli's role
in the evolution of QCD.  By evolution, I mean not only the evolution of
the parton distributions, for which Guido is justly famous, but also
the evolution in our ability to calculate with the QCD Lagrangian.
In the Autumn of 1972 I arrived in Rome as a
second-year graduate student, having been granted leave of absence
from the University of Oxford where I was enrolled in the Theoretical
Physics Department.  My motivations for leaving Oxford were not
entirely scientific, (see~\cite{Peierls}), but Rome La Sapienza turned
out to be a wonderful department, with an astonishing array of
talent. As well as Guido Altarelli, there were Franco Buccella, Nicola
Cabibbo, Raoul Gatto, Giorgio Parisi, Giuliano Preparata, and Massimo
Testa and fellow students Roberto Petronzio and later Guido
Martinelli.  The nearby institutions hosted Sergio Ferrara, Etim Etim,
Mario Greco, Luciano Maiani, Giulia Panchieri and Bruno Touschek.

As always in Italy, the organisational details were a little fuzzy and
real progress was only to be made by exploiting personal relationships. I
was formally assigned to study with Raoul Gatto, but the only office
available for me was the office of Giuliano Preparata in a small
corridor next to the office of Guido. I spent much of the first year
in Italy, learning Italian and adapting to the new way of life. I had
tried to work on a topic offered by Giuliano Preparata, but made
little progress and when Giuliano left to take up a position at CERN,
I started to work on a project suggested by Guido and Luciano
Maiani. Thus began a 12-year collaboration with Guido. He was my
mentor, collaborator and friend. After I moved to Fermilab in 1984, we
remained close friends until his untimely demise in 2015, although we
never collaborated again.

This intense period of our
relationship corresponded with the establishment of QCD, initiated by
the discovery of asymptotic freedom.  How QCD evolved from a
Lagrangian with the property of asymptotic freedom to a sophisticated
tool for the calculation of high-energy processes is the subject of
this note. I would like to identify five periods of great change
punctuating the evolution of the theory of the strong interactions,  
called Quantum Chromodynamics.
\begin{itemize}
\item 1970-1972, the pre-QCD years.
\item 1973-1974, the discovery of asymptotic freedom and the first applications.
\item 1976-1977, the Altarelli-Parisi equation, the demise of the $k_T$-cutoff, factorization 
and infra-red safety.
\item 1979, the Drell-Yan mechanism and the beginning of QCD corrections to hadronic processes,
and factorization beyond the leading logs.
\item 1983-1984, the discovery of the $W$ and the $Z$ and the conclusion of the S$p \bar{p}$S era.
\end{itemize}

Guido himself has written his own perspective on these
years~\cite{Altarelli:2011zv}.  Like that document, this note is not a
professional history, but rather a sketch of these years as I remember
them.
\section{Pre-QCD.}

The years 1973-1974 were watershed years for particle physics. Prior
to 1973 there were models of strong interactions, but no real
candidate theory.  Although approximate scaling was established in
deep inelastic scattering, the explanations for it were somewhat
baroque. One explanation was that the commutator of the
electromagnetic currents comported itself as a free field theory on
the light cone, but somehow conspired to be strongly interacting off
the light cone.  Another explanation was the parton
model~\cite{Feynman:1969ej,Bjorken:1969ja}, which required an 
{\it ad hoc} cutoff on the transverse momentum of the partons to explain the
scaling behaviour. The state of the theory was nicely summarized in
the book of Feynman~\cite{Feynman:1973xc}, written as a reaction to
the data presented at the Cornell conference~\cite{Cornell}.

Guido spent 1968-1969 at NYU and academic year 1969-1970 as a
Fulbright fellow at Rockefeller University.  Early in the spring of
1970, the group of Leon Lederman presented preliminary results on the
production of muon pairs at the meeting of the American Physical
Society. Two months later Drell and Yan produced their
paper~\cite{Drell:1970wh} proposing the quark antiquark annihilation
mechanism.
\begin{eqnarray}
\frac{d \sigma}{dQ^2} &=&\frac{4 \pi \alpha^2}{3 Q^2} \frac{1}{Q^2}  {\cal F}(\tau) \nonumber \\
&=& \frac{4 \pi \alpha^2}{3 Q^2} \frac{1}{Q^2} 
\int_0^1 dx_1 \int_0^1 dx_2 \delta(x_1 x_1 - \tau) \sum_{a} \lambda_a^{-2} F_{2a}(x_1)F^\prime_{2 \bar{a}}(x_2) 
\end{eqnarray}
$F_{2a}$ are (the components of $a$-type quarks and anti-quarks in) 
the deep inelastic structure functions, and $\lambda_a$ is the charge of the parton of type $a$. 
In a modern formulation there would be an additional
factor of $1/3$, due to the fact that e.~g. a red quark can only annihilate with an anti-red quark,

The final data of the Lederman group~\cite{Christenson:1970um},
published in September 1970, showed a rapidly falling continuum spectrum in
the mass $Q$ of the muon pairs with a shoulder in the region $Q
\approx 3-4$~GeV. With the benefit of hindsight the shoulder can be
ascribed to the $J/\psi$ and $\psi^\prime$ observed at low mass resolution.

Altarelli, Brandt and Preparata (ABP) followed the results
of Lederman's group closely, (perhaps too closely) and 
in September 1970 published a paper, 
based on a quark model with a scattering cross section having Regge-like 
properties which produced a shoulder in the region of $3$~GeV.
The model of ABP predicted non-scaling behaviour,
\begin{equation}
\frac{d \sigma}{dQ^2} \sim  \Big[\frac{1}{Q^2} F_1(\tau) +F_2(\tau)\Big]
\end{equation}
Although, ultimately the model of Drell and Yan gave the correct description, 
the familiarity with the muon-pair production process would prove to be important
for future research. 
\section{Asymptotic freedom.}

The years 1973-1974 were years of great change for both theoretical
and experimental particle physics.  The papers indicating that
non-Abelian gauge theories were asymptotically free were published in
June 1973 by Politzer~\cite{Politzer:1973fx} and by Gross and
Wilczek~\cite{Gross:1973id}.  1974 started with the
prediction that charmed quarks should have masses less than
$5$~GeV~\cite{Gaillard:1974hs}, based on the cancellations inherent in
the GIM mechanism~\cite{Glashow:1970gm}.  Subsequently, in November
1974, the $J/\psi$ was
discovered~\cite{Aubert:1974js,Augustin:1974xw}. The ADONE accelerator
at Frascati was able to raise the energy to 3.1GeV to produce the
$J/\psi$~\cite{Bacci:1974za}, and private communications with the
experimenters indicated an observed forward backward asymmetry in the
muon pairs produced on resonance.  This apparent asymmetry gave rise
to two papers from the Rome group, the first~\cite{Altarelli:1974wm},  
interesting but incorrect, claiming the particle discovered was the
$Z$-boson,
the second~\cite{Altarelli:1975za}, (presumably) correct,
but by the time it appeared uninteresting (because the large asymmetry had gone away),
calculating the forward-backward asymmetry in the presence of a vector
resonance.

In January of 1975, Guido and other members of the Rome group returned to consider the
Drell-Yan process, with an explicit model of the parton
distributions~\cite{Altarelli:1975qw}, this time taking into account
the production of the $J/\psi$.

\subsection{The first applications}

The acceptance of QCD as the correct theory of the strong interactions was slow. 
For a time many papers (including ours)
began with tentative phrases, such as {\it The gauge theory of colored quarks
and gluons (QCD) is at present the best candidate for a theory of
the strong interactions}~\cite{Altarelli:1977kt}.  Many influential
people felt that when a correct theory was found, it should instantly
make sense out of a disparate range of experimental data. Thus, as
late as 1976--1977, Feynman and Field, proposed their black box model
to explain inclusive scattering data~\cite{Field:1976ve}.  This
despite the fact that reliable lowest order cross sections for
parton-parton scattering became available in
1977~\cite{Combridge:1977dm}.  The energies were not high enough for
the $1/p_T^4$-behaviour characteristic of Rutherford-like scattering
to reveal itself.

In addition to the insufficient energy, the other reasons for this
theoretical hesitation are not hard to understand, and were presented
by Guido in his write-up of the 1983 session of the Moriond
conference.  {\it `Although QCD essentially imposes itself as the only
  theory of the strong interactions within reach of the weapon arsenal
  of conventional quantum field theory, yet QCD is still the less
  established sector of the standard model. Testing QCD is in fact
  more difficult than testing the electroweak sector. In the latter
  domain perturbation theory can always be applied. Also the leptons
  and the weak gauge bosons are at the same time the fields in the
  lagrangian and the particles in our detectors. Instead QCD is a
  theory of quarks and gluons while only hadrons are
  observable. Moreover perturbation theory can only be applied in
  those particular domains of the strong interaction where approximate
  freedom, which is only asymptotic, can be
  reached'}~\cite{Altarelli:1984fq}.

But the excitement of having a candidate theory of the strong interactions
was lost on no-one.  The first applications were limited to
processes governed by the operator product expansion.  In addition to
deep inelastic scattering, which was treated shortly after the discovery of asymptotic 
freedom by the protagonists themselves~\cite{DeRujula:1974rv,Gross:1974cs},
Ken Wilson had
suggested~\cite{Wilson:1970ag} that the $\Delta I =1/2$ rule might be
explained by strong interaction effects, but now one had a theory with
which one could calculate.  The challenge was taken up by Altarelli
and Maiani~\cite{Altarelli:1974exa} in Italy and and Gaillard and
Lee~\cite{Gaillard:1974nj} in the United States.  It was found that
in the standard model, the strong interactions
did indeed give an enhancement of about $4-6$ in amplitude, too small
to explain the whole observed amplitude enhancement of 20.
My thesis project
was to calculate whether there was a similar pattern of enhancement in
parity violating processes in nuclei which proceed via the weak
interaction~\cite{Altarelli:1974ni}. However at this point the real challenge was to
find a way of going beyond the operator product expansion.

\section{The Altarelli-Parisi equation}

The precursors of the paper on the Altarelli-Parisi equation were
presented at two back-to-back winter conferences at Flaine in the
French Alps in 1976 by Altarelli~\cite{Altarelli:1976dp} and by
Parisi~\cite{Parisi:1976qj}.  The first paper deals with the
translation of deep inelastic data, especially neutrino data, from the
language of the operator product expansion into scale dependent parton
distributions.  The second paper contains an early form of the AP
equation; versions of the splitting functions are presented, but
some of the details are wrong. I am acknowledged in the paper by
Parisi, but only because I helped to correct the English.
 
In September 1976, frustrated with the progress of his career in Rome,
Guido went on sabbatical to ENS in Paris. As luck would have it
Giorgio Parisi was also visiting Paris. The paper on the evolution of
the parton distributions was written there.  The Altarelli-Parisi
(DGLAP) equation~\cite{Altarelli:1977zs,Dokshitzer:1977sg} changed the
way that we thought about deep inelastic scattering. It made it quite
clear that the scale dependence of the parton distributions was
processs independent. In the paper the splitting functions were
calculated from the branching probabilities, without reference to any
particular hard scattering process.  This simple branching picture
only holds in a physical gauge in which only the transverse degrees of
freedom of the gluon field propagate. The splitting functions were
calculated in old-fashioned perturbation theory, in which manifest
Lorentz invariance is lost, but unitarity is simpler. Importantly, the
calculation thrust the attention back on the Feynman diagrams, rather
than the operator product expansion, the proofs of which were not
widely understood. Before the publication of their paper there was
little understanding of the diagrams resummed by the renormalization
group in the operator product treatment of deep inelastic scattering.

Guido worked on the manuscript of the paper~\cite{Altarelli:1977zs}
through the winter of 1976-1977, and I passed through Paris and was
shown a copy of the manuscript. He was concerned that the paper would
not be accepted for publication, because although the paper introduced
a new language for the unpolarized structure functions, the results
were simply the inverse Mellin transform of results previously
obtained by Christ, Hasslacher and Mueller~\cite{Christ:1972ms} for
the Abelian case, and Gross and Wilczek~\cite{Gross:1974cs} and Georgi
and Politzer~\cite{Georgi:1951sr} for the non-Abelian case.  To be
sure it was an inverse Mellin transform that required the introduction
of a new quantity, the plus distribution, similar in spirit to the
delta function of Dirac, but neverthless it was just an inverse Mellin
transform. To forestall any possible objection by a referee,
calculations of the polarized splitting functions, which had recently
been calculated by others~\cite{Ahmed:1976ye,Sasaki:1975hk} were included.

The treatment of Altarelli and Parisi raises the question of how QCD
and the parton model could be reconciled for the Drell-Yan process.
In the summer of 1977 Politzer wrote his first paper on
factorization~\cite{Politzer:1977fi} in hard processes treating the
specific case of muon-pair production.

\section{Drell Yan and the $K$-factor.}

The first order of business was to determine whether the transverse
momentum of the muon pairs, was limited as predicted by the naive
parton model, or whether it grew with $Q^2$ at fixed $Q/\sqrt{S}$ as
required by QCD. The issue was clouded by the fact that an experiment
running at fixed $\sqrt{S}$, could not directly investigate this scaling.
Detailed calculations~\cite{Altarelli:1977kt,Altarelli:1978pn} showed
that the average value of the transverse momentum was expected to be
approximately constant at fixed $\sqrt{S}$ and increasing $Q$, 
as the perturbative growth with $Q$ was balanced
against the fall-off of the structure functions.

The next challenge was to calculate higher order corrections to the
Drell-Yan process. From an operational point of view, parton
distributions measured in leptoproduction were used to predict cross
sections for muon pair production. In our
theoretical approach~\cite{Altarelli:1978id} we followed a similar logic, 
finessing the issue of how much was factorized into the parton distribution by
comparing the perturbative results for deep inelastic and Drell-Yan.
Thus parton distributions were defined beyond the leading order in terms
of the radiatively corrected deep inelastic structure function $F_2$.
As a consequence of the Adler sum rule, which has no perturbative corrections,
this choice had the nice feature that the number of valence quarks in the
proton remains fixed at all orders in QCD perturbation theory.
To perform these calculations a regulator is required to control the
divergences associated with soft and collinear parton emission. In our
first paper this was achieved by taking the quarks slightly off their
mass shell. This was doable, but quite cumbersome.  The second
paper~\cite{Altarelli:1979ub} regulated the singularities using
dimensional regularization, which was much more efficient. This is now
a textbook calculation, and established the method used to calculate
all higher radiative corrections to hard processes in QCD.  The
corrections to the Drell-Yan turn out to be large at the values of $Q$
which were probed at that time. In fact they were so large that one might doubt the
validity of perturbation theory.  The data~\cite{Badier:1979jd} also
showed an excess over the tree level prediction of the Drell-Yan
model by a factor of about two. This was dubbed the $K$-factor by Guido~\cite{Altarelli:1979zh}.  
Particularly significant was the data with antiproton beams,
which in the Drell-Yan picture proceeded via the annihilation of two valence quarks.
The occurrence of the same $K$-factor in this case
showed that ignorance of the true size of the sea quark
distributions, could not be responsible for the effect.

In 1979 I moved from MIT to Caltech. On arrival I was pleased to
discover that Feynman and Field were repeating the calculations that
we had done and published~\cite{Altarelli:1979ub,Altarelli:1979kv}
the previous year. The fact that Feynman
felt that our calculations were important enough to repeat was a big
psychological boost for me; Guido's leadership had placed us in the
major league.

\section{S$p\bar{p}$S collider: the discovery of the $W$ and $Z$.}

The S$p\bar{p}$S collider was an opportunistic project to exploit quark antiquark annihilation,
with the explicit goal of discovering the $W$ and the $Z$. 
Given our history with the Drell-Yan process we were perfectly poised to 
exploit the physics of the S$p\bar{p}$S collider, which in the main was the physics of 
quark-antiquark annihilation. 
To be sure, we were able to write authoritative 
papers on the total cross section for $W$ and $Z$ production, but also for the 
the $p_T$ and rapidity distributions for the produced bosons~\cite{Altarelli:1984pt,Altarelli:1984kp}.
Our theoretical prediction for the $p_T$ distribution of the $W$-bosons 
is quoted in the Nobel lecture of Carlo Rubbia~\cite{Fraengsmyr:1993ii}.

As an historical aside, I note that the discovery 
of the $W$ and $Z$ caused consternation in America,
since the bold strategy of building the S$p \bar{p}$S at CERN initiated the transfer
of the leadership in experimental high energy physics to Europe.  
This loss of primacy was especially bitter at Fermilab
since the proton anti-proton machine had been originally
proposed for the Fermilab main ring~\cite{Cline:1976cp}.
Although it might be a {\it post-facto} rationalization, 
the experiment would probably not have worked at Fermilab.    
``The vacuum system was terrible, so beam lifetime (number of
antiprotons) and, worse, luminosity lifetime (beam size due to
multiple scattering) would have been very poor. At injection (8 GeV)
the beam lifetimes were measured in seconds, which was made worse by poor
magnetic field at injection. As one accelerates those problems become
less severe, but the top energy was limited by the magnet design. 
The Main Ring magnets could not be operated DC at more than, probably, 200
GeV because they would burn up''~\cite{Limon}. 

After the great success of the S$p\bar{p}$S collider in discovering the 
$W$ and $Z$ bosons and confirming the standard model, 
there was a desire to exploit the machine to the utmost. Could it also give signatures 
of physics beyond the standard model? 
At the Bern Conference in 1984, Carlo Rubbia presented evidence for five events with greater than 
40 GeV of missing energy~\cite{Arnison:1984qu}, and, in my recollection, declared the Standard Model to be dead.   

The following year, directly after the presentation of Rubbia at the Saint-Vincent conference~\cite{Rubbia:1985ng},
Guido stated his opinion that a cocktail of standard model processes plus a few cracks
in the detector could explain the monojet events.  
Our theoretical understanding of the $p_T$ distributions of produced $W,Z$ bosons, gave us
great confidence in the predicted rates for monojets coming from $Z$+jet events,
with the $Z$ decaying to neutrinos. As stated by Roy Schwitters in his conference summary~\cite{Schwitters:1985nb},
`The basic point is that by a combination of improbable but conventional processes 
such as $Z^0$ + gluon where the $Z^0$ decays to neutrinos, single $W$ production followed by tau decay, 
and measurement errors, one may be eventually able to explain all the mono-jets. At the workshop 
this became known as the Altarelli cocktail.'
The Altarelli cocktail was the beginning of the end for beyond-the-standard-model
explanations of the monojet events.

\section{The man himself}
Guido was a masterful conference summarizer, managing to take stock of
all the important issues with great clarity. His presentations were
colourful and presented in an idiosyncratic English that left no doubt
about his Mediterranean origins.  I remember him with his tall frame
leaning slightly toward his interlocutor, his long elegant hands
gesticulating to emphasize his point.

As Guido's student I was given various pieces of advice.  One was to
cultivate `il senso del gioco', an understanding of the strategy of
the game, a concept taken from professional soccer.  Throughout his
professional life Guido displayed a consummate understanding of the
game, a keen understanding of what was important, and what was not.  
I may not have mastered `il senso del gioco' but I certainly learnt
from Guido's sense of fun, his wry sense of humour and his playful use
of the Italian language.

When my time as a post-doc in Rome came to an end, it was time to get 
letters of reference. Nicola Cabibbo was the most famous person in the department
with whom I had been in contact. So I summoned up my courage and asked him for 
a letter of recommendation. Nicola's answer was that if I wrote the letter of  
recommendation, he would sign it! (Nicola was not a man to trouble himself 
with things that did not interest him). I tried to draft such a letter, but found out
impossible. When informed of my dilemma, Guido's response was immediate and predictable,
`Write a letter which says that Keith Ellis is more talented than Nicola Cabibbo, and see if 
he signs it.'
\section{In conclusion}
It has been one of the most rewarding periods of my professional life
to work with Guido Altarelli.  With his leadership, we were able to
play our part in turning the fledgling theory of QCD into the sophisticated
calculational engine that it has become today. As a recapitulation of what was
achieved in those early years, I can do no better than quote Guido himself, and so here is
a sentence from his plenary talk~\cite{Altarelli:1985az} at the Bari
conference in 1985. 
`{\it The beautiful ‘naive’ parton model of Bjorken, Feynman and
  others has by now evolved into the ‘QCD improved’ parton model. This
  powerful language has become such a familiar and widespread tool for
  everyday practice in high energy physics that one is led to take all
  its new successes as granted and in a way obvious.}'
\bibliographystyle{varenna} \bibliography{GA_RKE.bib}
\end{document}